\documentclass[10pt]{iopart}
\usepackage{graphicx,iopams,cite}

\usepackage[T1]{fontenc}
\usepackage[latin1]{inputenc}

\newcommand{\eq}[1]{Eq.(\ref{#1})}
\newcommand{\fig}[1]{Fig.~\ref{#1}}
\newcommand{\sect}[1]{Section~\ref{#1}}

\newcommand{\eqd}[2]{Eqs.(\ref{#1}) and~(\ref{#2})}

\newcommand{\ahum}[1]{``#1''}
\newcommand{\avg}[1]{\langle#1\rangle}
\newcommand{\qavg}[1]{[#1]}

\newcommand{\LV}{liquid-vapor }

\newcommand{\tch}{T_{\rm c}}
\newcommand{\chidis}{\chi_{\rm dis}}
\newcommand{\pli}{P_{L,i}(s)}
\newcommand{\kb}{k_{\rm B}}

\newcommand{\sigmac}{\sigma_{\rm c}}
\newcommand{\sigmap}{\sigma_{\rm p}}
\newcommand{\sigmaq}{\sigma_{\rm Q}}

\newcommand{\etapr}{\eta_{\rm p}^{\rm r}}
\newcommand{\etaprcr}{\eta_{\rm p,cr}^{\rm r}}

\newcommand{\etac}{\eta_{\rm c}}
\newcommand{\mucx}{\mu_{\rm coex}}
\newcommand{\mucr}{\mu_{\rm cr}}

\begin{document}


\title[Colloid-polymer mixtures in random porous media]{Colloid-polymer mixtures 
in random porous media: Finite size scaling and connected versus disconnected 
susceptibilities}


\author{R. L. C. Vink$^{1,2}$, K. Binder$^3$, and H. Löwen$^2$}

\address{$^1$Institute of Theoretical Physics, Georg-August-Universität 
Göttingen, Friedrich-Hund-Platz~1, 37077 Göttingen, Germany}

\address{$^2$Institut für Theoretische Physik II: Weiche Materie,
Heinrich-Heine-Universität Düsseldorf, Universitätsstraße 1, 40225
Düsseldorf, Germany}

\address{$^3$Institut für Physik, Johannes-Gutenberg-Universität Mainz, 
Staudinger Weg 7, 55099 Mainz, Germany}

\begin{abstract} As a generic model for \LV type transitions in random porous 
media, the Asakura-Oosawa model for colloid-polymer mixtures is studied in a 
matrix of quenched spheres using extensive Monte Carlo (MC) simulations. Since 
such systems at criticality, as well as in the two-phase region, exhibit lack of 
self-averaging, the analysis of MC data via finite size scaling requires special 
care. After presenting the necessary theoretical background and the resulting 
subtleties of finite size scaling in random-field Ising-type systems, we present 
data on the order parameter distribution (and its moments) as a function of 
colloid and polymer fugacities for a broad range of system sizes, and for many 
(thousands) realizations of the porous medium. Special attention is paid to the 
\ahum{connected} and \ahum{disconnected} susceptibilities, and their respective 
critical behavior. We show that both susceptibilities diverge at the critical 
point, and we demonstrate that this is compatible with the predicted scenario of 
random-field Ising universality. \end{abstract}



\maketitle

\section{Introduction}

Understanding the behavior of fluids that undergo a \LV phase transition in the 
bulk (or, equivalently, of binary mixtures undergoing bulk phase separation), is 
still rudimentary when one considers the {\it confinement} of such systems in 
mesoporous materials, such as porous glasses or silica gels 
\cite{lev-d-gelb.gubbins.ea:1999}. Such amorphous materials form a highly 
irregular, interconnected, three-dimensionally percolating network, and \LV type 
transitions of fluids confined between the walls of such networks are widely 
observed and of practical importance \cite{gregg:1982}. However, the precise 
nature of the \LV critical point of such systems is still only partly understood 
\cite{wong.chan:1990, wong.kim.ea:1993, zhuang.casielles.ea:1996, 
tulimieri.yoon.ea:1999, gennes:1984, kierlik.rosinberg.ea:1996, 
alvarez.levesque.ea:1999, page.monson:1996, sarkisov.monson:2000, 
scholl-paschinger.levesque.ea:2001, schmidt.scholl-paschinger.ea:2002, 
kierlik.monson.ea:2002}. While de~Gennes \cite{gennes:1984} has presented a 
simple argument that the critical behavior of fluids in such random media can be 
mapped onto the random-field Ising model (RFIM) \cite{imry.ma:1975, imbrie:1984, 
nattermann:1998}, this prediction could be confirmed neither by experiments 
\cite{wong.chan:1990, wong.kim.ea:1993, zhuang.casielles.ea:1996, 
tulimieri.yoon.ea:1999}, nor by numerical calculations (MC simulations 
\cite{kierlik.rosinberg.ea:1996, alvarez.levesque.ea:1999, 
sarkisov.monson:2000}, or density functional theories 
\cite{scholl-paschinger.levesque.ea:2001, schmidt.scholl-paschinger.ea:2002, 
kierlik.monson.ea:2002}, respectively).

On the other hand, these studies could not point out any flaw in this more than 
twenty year old argument of de~Gennes \cite{gennes:1984} either. In short, the 
argument of de~Gennes starts from the well-known phenomenon of capillary 
condensation \cite{lev-d-gelb.gubbins.ea:1999, evans:1990, 
binder.landau.ea:2003}. In an infinitely long slit-pore, the \LV transition is 
\ahum{shifted} relative to the bulk, due to the attractive forces between the 
fluid particles and the walls. The magnitude of the shift depends on the nature 
of the fluid and the type of walls. In addition, if \ahum{drying} rather than 
\ahum{wetting} would occur for very thick slits, also the opposite effect of 
\ahum{capillary evaporation} may take place \cite{lev-d-gelb.gubbins.ea:1999, 
evans:1990, binder.landau.ea:2003}. In general, the chemical potential 
$\mucx(D)$ where liquid and vapor coexist inside a slit pore, differs from the 
bulk coexistence chemical potential $\mucx(\infty)$, and this difference depends 
on the width $D$ of the slit pore. In an irregular interconnected pore network, 
the local pore diameter at position $\vec{r}$ fluctuates randomly around some 
average value. As a result, the local chemical potential $\mu(\vec{r})$ where 
phase coexistence would occur, will also exhibit (quenched) fluctuations around 
some average value (the fluctuations are quenched because the structure of the 
porous network does not change over time). The analogy to the RFIM is readily 
seen when the fluid is described as a lattice gas, since the latter is 
isomorphic to the Ising ferromagnet. In terms of the Ising ferromagnet, quenched 
random fluctuations in the local chemical potential, become isomorphic to 
quenched external field variables $h_i$, with $h_i$ a random variable acting on 
the spin at the $i$-th lattice site (which is precisely the random-field Ising 
model). This reasoning is also easily carried over to binary fluid mixtures 
\cite{gennes:1984}.

In the present work, we contribute to the clarification of this problem, by 
presenting extensive MC data for a particularly simple model, namely the 
Asakura-Oosawa (AO) model \cite{asakura.oosawa:1954, vrij:1976} of 
colloid-polymer mixtures, inside a quenched random porous medium. The AO model 
is known to capture {\it bulk} experimental observations very well (by bulk we 
mean in the absence of any porous medium), including phase separation 
\cite{lekkerkerker.poon.ea:1992, aarts.tuinier.ea:2002} and interfacial 
properties \cite{brader.evans.ea:2002}. Computer simulations 
\cite{vink.horbach:2004*1, vink.horbach:2004, vink.horbach.ea:2004} have shown 
that bulk phase separation in the AO model, which occurs for sufficiently large 
polymers at sufficiently high polymer fugacity, belongs to the universality 
class of the Ising model \cite{fisher:1974, fisher.zinn:1998, zinn-justin:2001, 
binder.luijten:2001}. In addition, the standard predictions for capillary 
condensation in slit pores \cite{evans:1990, binder.landau.ea:2003, 
fisher.nakanishi:1981} have been well confirmed for this model 
\cite{vink:056118, vink:031601, virgiliis.vink.ea:2006, 
fortini.schmidt.ea:2006}. Following our previous work \cite{vink.binder.ea:2006, 
pellicane.vink.ea:2008}, we now consider the more complex problem of the AO 
model inside a random porous medium. The porous medium is obtained using an 
\ahum{easy} recipe: we simply distribute a set of obstacles (spheres) at random 
positions in the simulation box. Once the spheres have been positioned, they 
remain \ahum{fixed}, to mimic the quenched nature of the medium. Next, the AO 
model is inserted into the medium, and its phase behavior is studied. In 
particular, we will focus on (appropriately constructed) \ahum{susceptibilities} 
of the form $\qavg{\avg{\cdot}^2} - \qavg{\avg{\cdot}}^2$, with $\avg{\cdot}$ 
the conventional Gibbs-Boltzmann thermal average, and $\qavg{\cdot}$ an average 
over many different realizations of the quenched obstacles. Of course, in the 
{\it absence} of the porous medium, any such \ahum{susceptibility} is trivially 
zero. However, in its {\it presence}, the analogy to the random-field Ising 
model implies that such quantities will actually diverge at the critical point, 
and will do so with a characteristic critical exponent $\bar{\gamma}$.

The outline of our paper is as follows. In \sect{theory}, we recall in detail 
the necessary background of finite size scaling in the Ising and the 
random-field Ising models, and we discuss how these techniques may be 
carried-over to fluids with quenched disorder. Next, in \sect{method}, we define 
the AO model, explain how this model may be extended to also capture quenched 
disorder, and we describe our simulation method. The results are presented in 
\sect{result}, and we end with a discussion, conclusion, and summary in the last 
section.

\section{Finite size scaling in the Ising model, the random-field Ising model, 
and related models}\label{theory}

\subsection{Ising model}

We first consider the pure Ising model, i.e.~without any random field, and 
discuss how finite size scaling can be used to extract the critical properties 
of this model. To be specific, we consider a (nearest-neighbor) Ising 
ferromagnet on a hypercubic $d$-dimensional lattice of linear dimension $L$ and 
periodic boundary conditions
\begin{equation}
 {\cal H}_{\rm Ising} = -J \sum_{\avg{i,j}} s_i s_j
 - H \sum_i s_i, \hspace{1cm} s_i = \pm 1,
\end{equation}
with $J$ the exchange constant and $H$ an uniform external magnetic field. 
Defining the instantaneous magnetization per spin $s$ as
\begin{equation}
 \label{eq:spin}
 s = \frac{1}{L^d} \sum_i s_i,
\end{equation}
the object of interest is essentially the distribution
\begin{equation}
 \label{eq:pls}
 P_L(s) \equiv P_L(s|T,H),
\end{equation}
defined as the probability to observe a magnetization per spin $s$, in a system 
of size $L$, at temperature $T$ and field strength $H$. Basic observables of 
interest follow from the moments $\avg{s^k} \equiv \int_{-\infty}^{+\infty} s^k 
P_L(s) \, ds$ of the distribution, where $\avg{\cdot}$ is a conventional 
Gibbs-Boltzmann thermal average. For instance, the average magnetization per 
spin can be written as
\numparts
\begin{equation}
 \label{eq:mth}
 m(T,H) = \avg{s},
\end{equation}
and for the susceptibility $\chi$ we obtain
\begin{equation}
 \chi(T,H) = \frac{\partial m}{\partial H} = 
 L^d \left( \avg{s^2} - \avg{s}^2 \right), 
\end{equation}
\endnumparts
where the factor $\kb T$ has been absorbed in the definition of $\chi$, with 
$\kb$ the Boltzmann constant. We emphasize that these expressions must be used 
with care when $P_L(s)$ is bimodal. This happens, for example, at low 
temperature and $H=0$, since then a spontaneous magnetization exists, which may 
be positive or negative. Consequently, $P_L(s)$ has two peaks, one at positive 
and one at negative values. However, blindly applying \eq{eq:mth}, one 
finds that $m(T,0)=0$ irrespective of $T$, which is not really desirable. Since 
the Ising model has spin reversal symmetry, we have $P_L(-s) = P_L(+s)$, and so 
an easy fix is to introduce
\numparts
\begin{eqnarray}
 m'(T,H) = \avg{|s|}, \\
 \chi'(T,H) = L^d \left( \avg{s^2} - \avg{|s|}^2 \right),
\end{eqnarray}
\endnumparts
which are to replace $m$ and $\chi$ in these cases. The absolute value has the 
same effect as using a modified distribution $\left(P_L(s) + P_L(-s)\right)/2$ 
with the integration domain restricted to $s>0$. Clearly, such a modification is 
reasonable when $P_L(-s) = P_L(+s)$ somewhat holds. For very asymmetric 
distributions, a safer approach is to define $m$ and $\chi$ in terms of peak 
positions and widths, respectively. This approach was successfully applied to 
the AO model in \cite{vink.horbach.ea:2004} and will also be used in this work 
later on. Of course, in the thermodynamic limit, all definitions become 
equivalent, see discussion in~\cite{binder.heermann:2002}.

As is well known, for $d \geq 2$, the Ising model has a second order phase 
transition from the (disordered) high-temperature paramagnetic phase, to the 
(ordered) low-temperature ferromagnetic phase, at some critical temperature 
$\tch$. In the vicinity of $\tch$, we expect power law singularities 
\cite{fisher:1974}
\numparts
\begin{eqnarray}
 \label{eq:pl_mag}
 m(T,0) \propto (-t)^\beta \hspace{5mm} \mbox{(order parameter)}, \\
 \label{eq:pl_chi}
 \chi(T,0) \propto |t|^{-\gamma}, \\
 \label{eq:pl_xi}
 \xi \propto |t|^{-\nu},
\end{eqnarray}
\endnumparts
with $t \equiv T/\tch-1$ the reduced distance from the critical point, and $\xi$ 
the correlation length of the magnetization fluctuations. In the above, $\beta$, 
$\gamma$, and $\nu$ are critical exponents, which characterize the universality 
class. 

Of course, the divergence of the correlation length cannot be captured in a 
finite simulation box of size $L$, and so the above power laws are never 
observed directly. The state-of-the-art is to perform several simulations, using 
a range of system sizes $L$, and to extrapolate the simulation data to $L \to 
\infty$ via finite size scaling. In its simplest form, finite size scaling is 
just the statement that, in a finite system at the critical point $\xi 
\propto L$ \cite{fisher:1974}. Eliminating $t$ from \eqd{eq:pl_mag}{eq:pl_xi}, 
and using $\xi \propto L$, one immediately derives the $L$-dependence of the 
magnetization order parameter at the critical point
\numparts
\begin{equation}
 \label{eq:fss:op}
 m_L \propto L^{-\beta/\nu},
\end{equation}
Similarly, for the susceptibility, one obtains
\begin{equation}
 \label{eq:fss:chi}
 \chi_L \propto L^{\gamma/\nu}.
\end{equation}
\endnumparts
These equations simply state that, if one performs a simulation at the critical 
point over a range of system sizes, the magnetization should vanish $\propto 
L^{-\beta/\nu}$, and the susceptibility should increase $\propto 
L^{\gamma/\nu}$.

The above scaling laws are quite general, and should hold near any critical 
point where the correlation length diverges as a power law, i.e.~conform 
\eq{eq:pl_xi}. If also the hyperscaling relation is obeyed
\begin{equation}
 \label{eq:hs}
 \gamma + 2 \beta = \nu d,
\end{equation}
with $d$ the spatial dimension, it follows that the entire distribution $P_L(s)$ 
at $H=0$ scales with $L$ as \cite{binder:1981}
\begin{equation}
\label{eq:fss}
 \left. P_L(s) \right|_{H=0} = L^{\beta/\nu} \tilde{p}(L/\xi,s L^{\beta/\nu}),
\end{equation}
with $P_L(s)$ the magnetization distribution of \eq{eq:pls}. Here, 
$\tilde{p}(x,x')$ is a universal scaling function, which essentially depends on 
the universality class, and the scaling should hold in the limits $\xi \to 
\infty$, $L \to \infty$, with $L/\xi$ finite. Using \eq{eq:fss}, one readily 
obtains the moments
\numparts
\begin{eqnarray}
 \label{eq:mom1}
 \avg{|s|} &=& \int |s| P_L(s) \, ds = L^{-\beta/\nu} \tilde{f}_0(L/\xi), \\
 \label{eq:mom2}
 \avg{s^k} &=& \int s^k P_L(s) \, ds = L^{-k \beta/\nu} \tilde{f}_k(L/\xi)
 \hspace{5mm} (k>0), 
\end{eqnarray}
\endnumparts
which also define the scaling functions $\tilde{f}$. For $k=1$, one recovers the 
average magnetization of \eq{eq:mth}, and the expected scaling law 
\eq{eq:fss:op} is correctly reproduced. For the susceptibility, however, these 
moments imply $\chi \propto L^{d - 2\beta/\nu}$, consistent with \eq{eq:fss:chi} 
only when hyperscaling holds. For the Ising model, hyperscaling indeed holds 
\cite{fisher:1974}, and so the use of \eq{eq:fss} is justified here.

\begin{figure}
\begin{center}
\includegraphics[clip=,width=11cm]{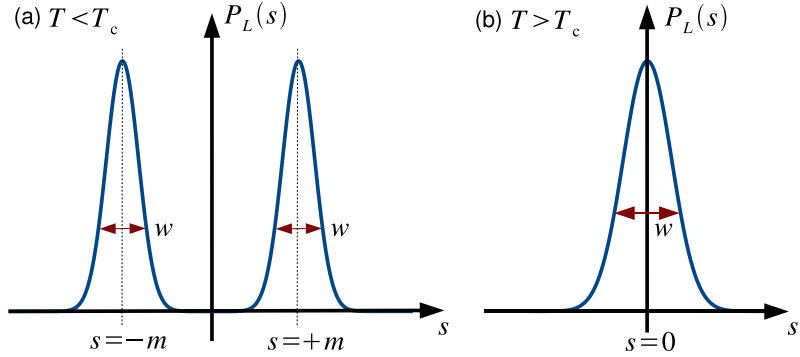} 
\caption{\label{plm} Schematic representations of the distribution $P_L(s)$ 
below the critical temperature $\tch$ (a) and above (b). Since the 
susceptibility for $T \neq \tch$ is finite, the peak widths $w$ vanish with 
increasing system size $\propto L^{-d/2}$, leading to a distribution 
featuring two $\delta$-peaks when $T<\tch$, and a single $\delta$-peak when $T> 
\tch$.}
\end{center}
\end{figure}

Hyperscaling also implies a remarkable property concerning the {\it shape} of 
$P_L(s)$ at criticality. To see this, note first that, below $\tch$, there 
exists a spontaneous magnetization. The magnetization may be positive ($+$) or 
negative ($-$), and so $P_L(s)$ features two peaks centered around $s = \pm m$, 
see \fig{plm}(a). Each of the peaks may be approximated by a Gaussian 
\cite{binder:1981}
\begin{equation}
 \label{eq:gauss}
 P_L^\pm(s) \approx L^{d/2} (2 \pi T \chi)^{-1/2} \exp \left[
 -(s \mp m)^2 L^d / (2 T \chi) \right],
\end{equation}
leading to a squared peak width $w^2 \equiv \avg{s^2} - \avg{s}^2 = \chi T / 
L^d$. In the thermodynamic limit, the peaks remain at their respective positions 
$s \pm m$. At the same time, since the susceptibility away from $\tch$ is 
finite, the peaks also become increasingly narrow, eventually converging to a 
distribution consisting of two $\delta$-peaks. The behavior in the disordered 
region $T > \tch$ follows analogously. In this case, a spontaneous magnetization 
is absent, and $P_L(s)$ is just a single Gaussian, centered around $s=0$, see 
\fig{plm}(b). For $T > \tch$, $P_L(s)$ thus converges to a single $\delta$-peak 
in the thermodynamic limit. Mathematically, the shape of the distribution in 
these two limiting cases can be expressed using the cumulant $U_1 = \avg{s^2} / 
\avg{|s|}^2$ \footnote{Of course, the cumulant is to be calculated for the full 
distribution. In particular, for $T<\tch$, one should write $P_L(s) = (P_L^- + 
P_L^+)/2$.}. After some algebra, one finds that
\numparts
\begin{eqnarray}
 \lim_{L \to \infty, T<\tch} U_1 = 1 \hspace{5mm} 
 \mbox{(two $\delta$-peaks)}, \\
 \lim_{L \to \infty, T>\tch} U_1 = \pi/2 \hspace{5mm} 
 \mbox{(one $\delta$-peak)}.
\end{eqnarray}
\endnumparts
Precisely at $\tch$, the behavior of the cumulant is more subtle 
\cite{binder:1981}. In a finite system at $\tch$, the magnetization vanishes 
$\propto L^{-\beta/\nu}$, see \eq{eq:fss:op}. Hence, $P_L(s)$ still exhibits two 
peaks, at positions $\propto \pm L^{-\beta/\nu}$. For the root-mean-square peak 
width, we obtain $w^2 \equiv \avg{s^2} - \avg{s}^2 \propto L^{\gamma/\nu - d}$, 
where now \eq{eq:fss:chi} was used. Comparing the {\it distance} between the 
peaks to their {\it widths}, we find
\begin{equation}
 \label{eq:pw}
 \Delta \equiv \frac{\mbox{peak width}}{\mbox{peak-to-peak distance}}
 \propto L^\omega,
\end{equation}
with $\omega = (\gamma/\nu - d)/2 + \beta/\nu$. By virtue of hyperscaling one 
has $\omega=0$, implying that the relative peak width $\Delta$ does not vanish 
in the thermodynamic limit. Consequently, $P_L(s)$ at criticality does {\it not} 
become a superposition of two $\delta$-functions, but instead converges to a 
distribution of two overlapping peaks. The cumulant $U_1^\star$ at $\tch$ 
differs therefore from the off-critical values \cite{binder:1981}. By using 
\eqd{eq:mom1}{eq:mom2}, $U_1^\star$ can be expressed in terms of the scaling 
functions as $U_1^\star = \tilde{f}_2(L/\xi) / \tilde{f}_0^2(L/\xi)$, which is a 
universal function of $L/\xi$, and tends to a universal finite constant. In 
simulations, this result is useful since plots of $U_1$ versus $T$ for various 
system sizes $L$ will show a common intersection point, yielding an estimate of 
both $U_1^\star$ and $\tch$ (cumulant intersection method \cite{binder:1981}).

\subsection{random-field Ising model: finite size scaling}
\label{sec:fss_rfim}

The analysis of MC simulation \cite{binder.heermann:2002, newman.barkema:1999} 
data for systems belonging to the universality class of the random-field Ising 
model (RFIM) \cite{nattermann:1998} has certain subtleties 
\cite{eichhorn.binder:1995, eichhorn.binder:1996}, when one tries to apply 
finite size scaling methods \cite{fisher:1974, binder:1981}. The source of the 
problem is that the standard hyperscaling relation \cite{fisher:1974} between 
critical exponents, which is required by \eq{eq:fss} \cite{binder:1981}, does 
not hold for the RFIM \cite{nattermann:1998, villain:1982, schwartz:1985, 
fisher:1986, schwartz.gofman.ea:1991}. Since the presence of the random field 
breaks the spin reversal symmetry, it is necessary to consider also the 
\ahum{disconnected} susceptibility $\chidis$ \cite{schwartz:1985}, in addition 
to the standard \ahum{connected} susceptibility~$\chi$.

The RFIM Hamiltonian reads as
\numparts \begin{equation} \label{eq:rfim}
 {\cal H}_{\rm RFIM} = -J \sum_{\avg{i,j}} s_i s_j 
 - H \sum_i s_i - \sum_i h_i s_i, \hspace{1cm} s_i = \pm 1,
\end{equation}
with $J$ and $H$ defined as before. In addition, at each lattice site $i$, there 
acts a quenched random field $h_i$, which we take to be completely uncorrelated 
between neighboring sites, and with an average of zero
\begin{equation}
 h_i = \pm h, \hspace{1cm} \qavg{h_i}=0.
\end{equation}
\endnumparts
The amplitude $h$ of the random field should be small but finite ($h/J \ll 1$), 
but we are not concerned with the crossover to the pure Ising model here, and 
hence disregard the limit $h \to 0$. Basic observables are again the average 
magnetization per spin~$m$, the connected susceptibility $\chi$, and the 
disconnected susceptibility $\chidis$
\numparts
\begin{eqnarray}
 \label{eq:op}
 m(T,H) = \qavg{\avg{s}}, \\
 \chi(T,H) = L^d \qavg{\avg{s^2} - \avg{s}^2}, \\
 \label{eq:chidis}
 \chidis(T,H) = L^d \qavg{\avg{s}^2}.
\end{eqnarray}
\endnumparts 
For the same reason as before, we also introduce
\begin{equation}
 m'(T,H) = \qavg{\avg{|s|}}, \hspace{5mm}
 \chi'(T,H)  = L^d \qavg{\avg{s^2} - \avg{|s|}^2}.
\end{equation}
For the RFIM model, one has to perform the standard Gibbs-Boltzmann thermal 
average $\avg{\cdot}$ for one realization of the random field, followed by an 
average over~$M$ different random field configurations $\qavg{\cdot}$, whereby 
$M$ should be large. Note that $\chidis$ is simply the fluctuation of the 
average magnetization $\avg{s}$ between different realizations of the random 
field. Due to random variations in these fields, $\avg{s}$ will sometimes be 
negative, and sometimes be positive. In the limit $M \to \infty$, one has 
$\qavg{\avg{s}}=0$, of course, but the fluctuation $\qavg{\avg{s}^2} - 
\qavg{\avg{s}}^2$ will generally {\it not} be zero, which is essentially what 
$\chidis$ corresponds to. We shall also be interested in the distributions
\begin{equation}
 \pli \equiv P_{L,i}(s|T,H_i), \hspace{1cm} (i=1,\ldots,M),
\end{equation}
defined as the probability to observe a magnetization per spin $s$, in a system 
of size $L$, at temperature $T$ and external field $H_i$, for the $i$-th random 
field realization. Note that we allow $H_i$ to vary between different random 
field realizations. Ideally, one would like to have $M \to \infty$, but since 
resources are limited, simulations always deal with finite $M$.

Assuming that the RFIM, for small enough $h$, has a second order phase 
transition at $\tch$, we expect power law singularities for $m$, $\chi$, and 
$\xi$ as before, but with different critical exponents characteristic of the 
RFIM universality class \cite{nattermann:1998, schwartz:1985}. In addition, a 
power law is expected for the disconnected susceptibility
\begin{equation}
 \chidis \propto |t|^{-\bar{\gamma}},
\end{equation}
with a new critical exponent $\bar{\gamma}$ \cite{nattermann:1998, 
schwartz:1985}. It has been proved rigorously that the RFIM in $d=3$ dimensions, 
at low enough temperature, indeed exhibits a nonzero spontaneous magnetization 
\cite{imbrie:1984}. It has not, however, been proved that the second order 
transition assumed above actually exists (also weak first order transitions 
\cite{young.nauenberg:1985}, or spin-glass type phases 
\cite{mezard.monasson:1994} have been suggested). Recent MC simulations, 
however, favor a second order transition, albeit that the critical exponents are 
still not known very accurately \cite{rieger:1995, newman.barkema:1996}.

While for the pure Ising model we have the standard hyperscaling relation 
between critical exponents \cite{fisher:1974}, for the RFIM, rather a different 
relation has been proposed \cite{villain:1982}
\begin{equation}
 \gamma + 2 \beta = \nu (d - \theta).
\end{equation}
Here, $\theta$ is an exponent which measures the deviation from the standard 
hyperscaling relation; when it is zero, standard hyperscaling is again 
recovered. Using the further result that $\theta = \gamma / \nu$ 
\cite{schwartz:1985}, it follows that $(\gamma + \beta)/\nu = d/2$.

We now discuss finite size scaling in the RFIM, following Eichhorn and Binder 
\cite{eichhorn.binder:1995, eichhorn.binder:1996}. Note first that the 
\ahum{derivation} of \eqd{eq:fss:op}{eq:fss:chi} still holds. Hence, $m$ and 
$\chi$ scale with $L$ as before, albeit with different exponents. Similarly, for 
the scaling of the disconnected susceptibility at $\tch$, we expect that
\begin{equation}
 \label{eq:fss_dis}
 \chi_{L,\rm dis} \propto L^{\bar{\gamma}/\nu}.
\end{equation}
If we assume that each distribution $\pli$ scales conform \eq{eq:fss}, it 
follows that $\avg{|s|}_i$ and $\avg{s^k}_i$ scale according to 
\eqd{eq:mom1}{eq:mom2}, respectively (the subscript denotes that the thermal 
average was taken in the $i$-th random field realization). Of course, the 
scaling functions $\tilde{f}$ may depend on the particular random-field 
realization, but the leading $L$ dependence will be the same each time. Since by 
definition $\qavg{\avg{X}} \equiv (1/M) \sum_{i=1}^M \avg{X}_i$, it follows 
trivially that the $L$-dependence implied by \eqd{eq:mom1}{eq:mom2}, appears in the 
quenched average also. We thus obtain
\begin{equation}
 \qavg{\avg{|s|}^2} = \hat{c}_0 L^{-2\beta/\nu}, \hspace{2mm}
 \qavg{\avg{s}^2} = \hat{c}_1 L^{-2\beta/\nu}, \hspace{2mm}
 \qavg{\avg{s^2}} = \hat{c}_2 L^{-2\beta/\nu}, \hspace{2mm}
\end{equation}
with redefined scaling functions $\hat{c}$, which can be expressed in terms of 
the functions $\tilde{f}$, of course, but for our subsequent discussion the 
precise form does not matter. Using the definitions of $\chi$, $\chi'$ and 
$\chidis$ the above equation implies
\begin{equation}
 \label{eq:morem}
 \hspace{-10mm}
 \chi =  \left(\hat{c}_2 - \hat{c}_1 \right) L^{d-2\beta/\nu}, \hspace{2mm}
 \chi' =  \left(\hat{c}_2 - \hat{c}_0 \right) L^{d-2\beta/\nu}, \hspace{2mm}
 \chidis =  \hat{c}_1 L^{d-2\beta/\nu}.
\end{equation}
On the other hand, finite size scaling also demands that $\chi \propto \chi' 
\propto L^{\gamma/\nu}$ and $\chi_{L,\rm dis} \propto L^{\bar{\gamma}/\nu}$. The 
solution of the paradox is to require that
\begin{equation}
 \label{eq:hsfake}
 \bar{\gamma} + 2\beta = \nu d,
\end{equation}
which correctly sets the scaling of $\chidis$, and also that $\hat{c}_0 = 
\hat{c}_2$ and $\hat{c}_1 = \hat{c}_2$. Note that \eq{eq:hsfake} is just the 
standard hyperscaling relation, but with $\gamma$ replaced by~$\bar{\gamma}$. 
Hence, even though normal hyperscaling in the RFIM does not hold, \eq{eq:fss} 
still gives a consistent description of finite size scaling, but one must accept 
that the connected susceptibility is not described by it, since the leading 
terms in $\chi$ and $\chi'$ cancel. To also describe the scaling of $\chi$ and 
$\chi'$, one needs to include the leading correction to scaling. This correction 
can be derived by assuming that $\pli$ at and below $\tch$ is a superposition of 
two Gaussians. Expressing the peak at positive magnetization as $P_{L,i}^+(s) 
\propto \exp \left( -(s-m_i)^2 / (2 w_i^2) \right)$, it follows that 
$\avg{s}_i=m_i$ and $\avg{s^2}_i = m_i^2 + w_i^2$. Performing the quenched 
average, we now obtain a non-zero expression for the connected susceptibility 
$\chi = (L^d/M) \sum_{i=1}^M w_i^2$. This term, consequently, is the sought-for 
correction; finite size scaling then implies that $w_i^2 \propto 
L^{\gamma/\nu-d}$ at criticality.

Since $\hat{c}_0 = \hat{c}_2$, it also follows that the cumulant at criticality 
$U_1^\star \equiv \qavg{\avg{s^2}} / \qavg{\avg{|s|}^2}$ in the RFIM tends to 
unity \cite{eichhorn.binder:1995, eichhorn.binder:1996}. The {\it shape} of the 
quenched-averaged distribution {\it at} $\tch$ is therefore similar to that {\it 
below} $\tch$: both distributions are characterized by $U_1=1$ in the 
thermodynamic limit. Hence, also at $\tch$, we have a distribution featuring two 
$\delta$-peaks \footnote{Of course, whereas for $T=\tch$ the peak positions 
scale $\propto L^{-\beta/\nu}$, they saturate at finite values $\pm m$ when $T < 
\tch$.}. This is profoundly different from systems where hyperscaling holds, 
since here $U_1^\star$ tends to a non-trivial value different from the 
off-critical values (as explained in the previous section). For the RFIM, plots 
of $U_1$ versus $T$, for various system sizes $L \to \infty$, no longer 
intersect. In practice, however, the system sizes feasible in simulations are 
still quite small, and so one is plagued by \ahum{cross-over} effects 
\cite{freire.oconnor.ea:1994} (in this case from Ising to RFIM universality). 
This means that an intersection point can typically still be identified, but it 
occurs at a value much closer to $U_1^\star=1$ of the RFIM \cite{rieger:1995, 
eichhorn.binder:1996, vink.binder.ea:2006}~\footnote{Note in particular Figure~7 
of \cite{rieger:1995} for the RFIM.}. The fact that the quenched-averaged 
distribution in the RFIM remains sharp at criticality, in contrast to 
overlapping, is also obvious from \eq{eq:hsfake}. Considering again the ratio 
$\Delta$ between peak-width and peak-to-peak distance, i.e.~conform \eq{eq:pw}, 
one finds that $\omega = (\gamma - \bar{\gamma}) / (2\nu)$. Using the result of 
Schwartz that $\bar{\gamma} = 2\gamma$ \cite{schwartz:1985}, it immediately 
follows that $\omega<0$. In other words, for the RFIM at its critical point, the 
relative peak width $\Delta$ vanishes, leading to a distribution featuring two 
$\delta$-peaks.

\subsection{random-field Ising model: sample-to-sample fluctuations}

The result of Schwartz \cite{schwartz:1985}, namely that $\bar{\gamma} = 
2\gamma$, can be made plausible when we consider one particular realization of 
the random field. In a volume $L^d$, roughly half the lattice sites \ahum{feel} 
a negative random field (and the other half a positive random field, obviously) 
but with Poissonian fluctuations. Hence, there will typically be an excess 
Zeeman energy of order $\pm h L^{d/2}$, which has the same physical effect as if 
an uniform external field of strength
\begin{equation}
 \label{eq:hsc}
 H_{\rm c} \sim \pm h L^{-d/2},
\end{equation}
acted on the spins in this volume (recall that $h$ is the strength of the random 
field). But then we expect a non-zero magnetization $\avg{s} = H_{\rm c} \chi 
\sim \chi h L^{-d/2}$ in this sample, with $\chi$ the connected susceptibility. 
Using near $\tch$ the standard finite size scaling relations for $\avg{s}$ and 
$\chi$, we obtain $L^{-\beta/\nu} \propto L^{\gamma/\nu-d/2}$, or $\beta/\nu = 
d/2 - \gamma/\nu$. Combining with \eq{eq:hsfake} one finds that $\bar{\gamma} = 
2\gamma$.

\begin{figure}
\begin{center}
\includegraphics[clip=,width=13cm]{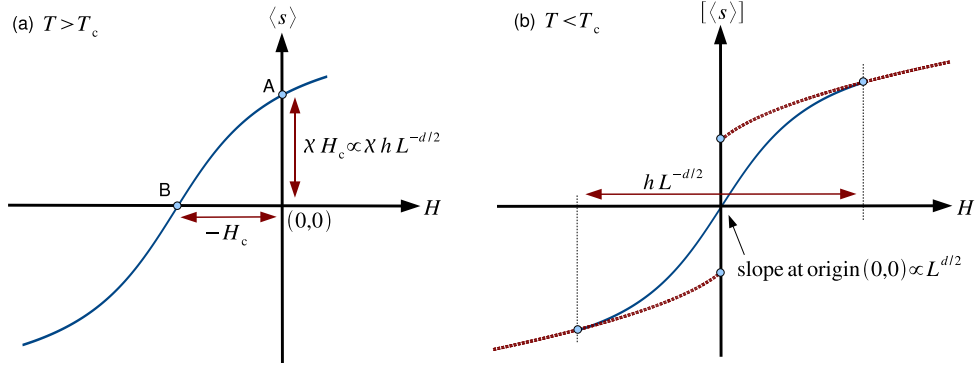}
\caption{\label{fig2} (a) $\avg{s}$ versus $H$ at $T>\tch$, for one 
realization of the random field where $H_{\rm c}$ happens to be positive. At 
$H=0$, we have a finite magnetization of order $\chi h L^{-d/2}$ (point A). The 
field at which $\avg{s}$ changes sign, and where $\chi \propto \partial \avg{s} 
/ \partial H$ attains its maximum, occurs at $H=-H_{\rm c}$ (point B). Note that 
the slope at $B$ approaches the zero-field connected susceptibility $\left. \chi 
\right|_{H=0}$ in the limit $L \to \infty$. (b) $\qavg{\avg{s}}$ versus $H$ at 
$T<\tch$. The dashed curve shows the behavior in the thermodynamic limit; the 
solid curve in a finite system of size~$L$. Note that the rounding is of order 
$L^{-d/2}$. This means that the slope $\left. \partial \qavg{\avg{s}} / 
\partial H \right|_{H=0}$ in finite systems grows $\propto L^{d/2}$, and that 
the region where $\qavg{\avg{s}}$ deviates significantly from $L \to \infty$ 
behavior shrinks $\propto L^{-d/2}$.}
\end{center}
\end{figure}

It is of some interest to explore the consequences of \eq{eq:hsc} further, and 
study the behavior of the magnetization for different realizations of the random 
field. If $H_{\rm c} > 0$ and $T>\tch$, we have for $H=0$ a positive 
magnetization of order $\avg{s} \sim \chi h L^{-d/2}$ as argued above (recall 
that $H$ is the strength of the uniform external field). The field at which the 
susceptibility $\chi \propto \partial \avg{s} / \partial H$ is maximized is 
therefore not $H=0$, but rather $H=-H_{\rm c}$, where the net effect of the 
random field is canceled. This is sketched in \fig{fig2}(a), where $\avg{s}$ 
versus $H$ is plotted \footnote{Of course, the graph of $\qavg{\avg{s}}$ versus 
$H$ is anti-symmetric about the origin, since in the quenched average both signs 
of $H_{\rm c}$ appear equally often.}. Taking the limit $L \to \infty$, it 
follows that the {\it slope} of the curve at points~A and~B becomes the same, 
since, on the small scale of $H_{\rm c}$, the curve may be approximated by a 
straight line. Note that the slope approaches the zero-field susceptibility 
$\chi$, and also that the slope is independent of $H_{\rm c}$. Plotting 
$\avg{s}$ versus $H$ for different realizations of the random field, one thus 
obtains a set of parallel straight lines. In other words, $\chi$ is rather 
insensitive to the particular random field configuration, which just expresses 
the fact that the system is self-averaging for $T>\tch$, as expected 
\cite{wiseman.domany:1995, wiseman.domany:1998}. This result is important since, 
in systems lacking spin reversal symmetry, the natural path in the $(T,H)$-plane 
to follow is no longer the line $H=0$, but rather the path along which $\chi$ 
assumes its maximum for each realization of the random field.

\begin{figure}
\begin{center}
\includegraphics[clip=,width=13cm]{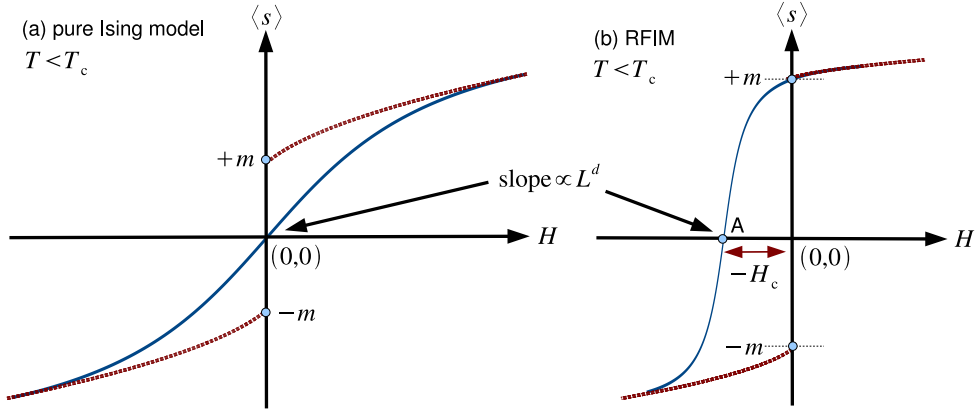}
\caption{\label{fig3} $\avg{s}$ versus $H$ at $T<\tch$ for the pure Ising model 
(a) and the RFIM (b). The dashed curves show the behavior in the thermodynamic 
limit; solid curves for finite systems (see details in text). Note that scenario 
(b) holds only in spatial dimension $d>2$.}
\end{center}
\end{figure}

The situation is qualitatively different for $T<\tch$, of course, since we now 
expect a first-order transition and, consequently, two-phase coexistence. In the 
pure Ising model, coexistence between two states (with positive and negative 
spontaneous magnetization $\pm m$) occurs at $H=0$, irrespective of the system 
size $L$. In the thermodynamic limit, $\avg{s}$ increases monotonically with 
$H$, and jumps from $-m$ to $+m$ at $H=0$; see the dotted curve in 
\fig{fig3}(a). In a finite system, the transition is rounded, and a true jump 
does not appear. Instead, $\avg{s}$ passes smoothly through the origin, but with 
slope $\partial \avg{s} / \partial H \propto L^d$ \cite{privman.fisher:1983, 
binder.landau:1984}, consistent with the formation of a jump in the 
thermodynamic limit; see the full curve in \fig{fig3}(a). Consequently, phase 
coexistence in the pure Ising model may always be studied using $H=0$. Provided 
$T$ is sufficiently below $\tch$, double-peaked distributions $P_L(s)$ are 
readily observed, i.e.~conform \fig{plm}(a), from which the coexistence 
properties follow.

For the RFIM below $\tch$ and finite system size $L$, the behavior is more 
subtle, since the random field breaks the spin reversal symmetry. We still 
expect a (rounded) first-order transition, but centered around the shifted field 
$H = - H_{\rm c}$, see \fig{fig3}(b). In the thermodynamic limit, $H_{\rm c} \to 
0$, and so the magnetization jumps, as before, at $H=0$ (dotted curve). In a 
finite system $\avg{s}$ increases smoothly with $H$ (solid curve), passing 
through zero at $H = - H_{\rm c}$ (point A), with slope $\partial \avg{s} / 
\partial H \propto L^d$. Since $H_{\rm c} \propto h L^{-d/2}$ asymptotically 
exceeds the rounding, it follows that, in a finite system at $H=0$, phase 
coexistence is unlikely. At $H=0$, one either observes the phase with positive 
magnetization (as one would in \fig{fig3}(b)), or, if the random field happens 
to resemble $H_{\rm c}<0$, a negative magnetization. Only very rarely, when the 
inflection point~A happens to coincide with $H=0$, will both phases be observed 
simultaneously. Hence, at $H=0$, the distribution $P_{L,i}(s)$ will mostly 
feature just one peak, located at positive or negative values. In the quenched 
average, one recovers $\qavg{\avg{s}}=0$, of course, but the fluctuation 
$\qavg{\avg{s}^2} - \qavg{\avg{s}}^2$ is not zero, since this, apart from a 
factor $L^d$, is precisely the disconnected susceptibility, see \eq{eq:chidis}. 
Clearly, to study phase coexistence in simulations, it does not make sense to 
use $H=0$, since one would rarely see a double-peaked distribution. Instead, it 
is more meaningful to obtain these properties at the inflection point~A, where 
$\chi \propto \partial \avg{s} / \partial H$ attains its maximum. To be precise: 
one should apply an external field $H = - H_{\rm c}$ \ahum{tailored} for each 
random-field realization. In the limit $L \to \infty$, one has $H_{\rm c} \to 
0$, and coexistence properties obtained at the inflection point, will agree with 
those obtained at $H=0$. The advantage of the former method being that 
double-peaked distributions $P_{L,i}(s)$ will now already appear in much smaller 
systems. Of course, for these double-peaked distributions, $\avg{s}_i$ will be 
close to zero each time, and so it follows that a different definition for the 
disconnected susceptibility should be used, presumably of the form $\chidis' = 
L^d \qavg{\avg{|s|}^2}$.

Considering now the behavior of $\qavg{\avg{s}}$ versus $H$ below $\tch$, we 
expect the scenario of \fig{fig2}(b). In the thermodynamic limit, a jump in 
$\qavg{\avg{s}}$ at $H=0$ is anticipated. In finite systems, the jump is 
rounded, but on a more severe scale $L^{-d/2}$, as pointed out by Kierlik \etal 
\cite{kierlik.monson.ea:2002}. Note that graphs of $\qavg{\avg{s}}$ versus $H$ 
for finite $L$ intersect the origin since, in the quenched average, both signs 
of $H_{\rm c}$ are equally likely.

Unfortunately, these arguments cannot be easily extended to $T=\tch$. As 
discussed in detail by Wiseman and Domany \cite{wiseman.domany:1995, 
wiseman.domany:1998}, systems with quenched random disorder at criticality 
exhibit lack of self-averaging. For small enough fields $H$, it still holds that 
$\avg{s}$ versus $H$ for one realization of the random field, is a straight 
line, with slope $\propto L^{\gamma/\nu}$. The same holds for $\qavg{\avg{s}}$ 
versus $H$, where the slope is also $\propto L^{\gamma/\nu}$, but the prefactors 
differ. The ratio of these prefactors is a quantity characterizing the lack of 
self-averaging, in the sense of Wiseman and Domany \cite{wiseman.domany:1995, 
wiseman.domany:1998}.

\subsection{obtaining the quenched average using a sample dependent $H_i$}

For the Ising model, one knows beforehand that the inflection point of $\avg{s}$ 
versus $H$ (or $\qavg{\avg{s}}$ versus $H$ in case of the RFIM), occurs on the 
symmetry line $H=0$. Hence, varying $T$ at fixed $H=0$, one cannot miss the 
critical point. In less symmetric models, the field $H$ at the inflection point 
is not known beforehand. In these cases, it is clearly more convenient to follow 
the path $H=-H_{\rm c}(T)$ in the $(T,H)$-plane of each random-field 
realization. That is, for each realization of the random field~$i$, one 
numerically locates the field $H_i$ where $\partial \avg{s}_i / \partial H$ in 
that sample is maximized. Properties of interest are then collected at $H_i$, 
and the process is repeated over many different random field samples. 
Extrapolating the data to $L \to \infty$ is demanding in practice, but does not 
present any principal objections. Only the {\it prefactors} of the finite size 
scaling laws at criticality
\begin{equation}
 \label{eq:ob}
 \overline{\avg{s}} \propto L^{-\beta/\nu}, \hspace{5mm}
 \overline{\avg{s^2} - \avg{s}^2} \propto L^{\gamma/\nu-d}, \hspace{5mm}
 \overline{\avg{s}^2} \propto L^{\bar{\gamma}/\nu-d},
\end{equation}
will differ from those of the standard quenched-average $\qavg{\cdot}$ obtained 
at fixed $H$. Here, the overbar denotes averaging at the sample dependent $H_i$. 
Since many typical fluids (including the AO model) are asymmetric, collecting 
the quenched average as $\overline{X}$ simply becomes a necessity in these 
cases.

\subsection{extension to fluids}
\label{sec:fluid}

We now consider a \LV transition of a fluid confined to a quenched porous 
medium. We use the grand canonical (GC) ensemble, i.e~volume~$L^d$, 
temperature~$T$, and chemical potential~$\mu$ are fixed, but the number of 
particles~$N$ in the system fluctuates. Our analysis is based on the 
(normalized) distribution
\begin{equation}
 \label{eq:pn}
 P_{L,i}(N) \equiv P_{L,i}(N|T,\mu_i), \hspace{1cm} (i=1,\ldots,M),
\end{equation}
defined as the probability to observe a system containing $N$ particles, in the 
$i$-th realization of the porous medium. Note the dependence on $L$ and $T$, and 
also that we allow the chemical potential $\mu_i$ to vary between different 
realizations of the porous medium. For given $L$ and $T$, $P_{L,i}(N)$ is 
sampled from $N=0$ to $N_{\rm max}$, using a biased sampling scheme 
\cite{virnau.muller:2004}. This process is repeated for $M$ different 
realizations of the porous medium. The sampling scheme is constructed to visit 
the full range $0 \leq N \leq N_{\rm max}$ irrespective of the imposed chemical 
potential. Hence, we set $\mu_i=0$ in the simulations, and use histogram 
reweighting \cite{ferrenberg.swendsen:1988} to extrapolate to different values 
afterward.

\begin{figure}
\begin{center}
\includegraphics[clip=,width=9cm]{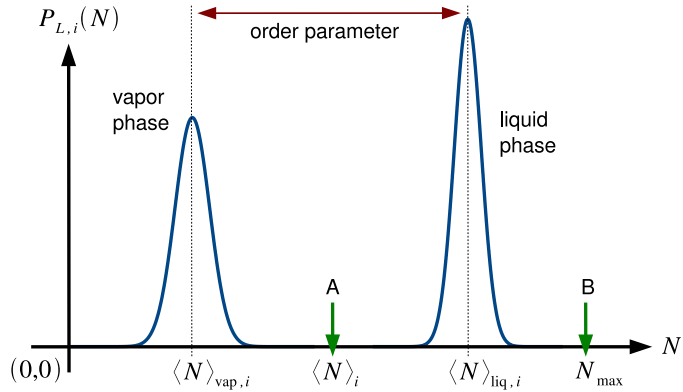}

\caption{\label{fig4} Schematic representation of the expected bimodal form of 
$P_{L,i}(N)$ obtained in a single realization of the porous medium at and below 
$\tch$. The average of the full distribution $\avg{N}_i$ separates the vapor 
from the liquid peak. The distance between the peaks, defined as the average 
number of particles in the liquid phase $\avg{N}_{{\rm liq},i}$ minus the 
average $\avg{N}_{{\rm vap},i}$ of the vapor, gives an estimate of the order 
parameter. The moments of the vapor and liquid peak yield the connected and 
disconnected susceptibilities, see details in text.}

\end{center}
\end{figure}

Assuming that the \LV transition in a porous medium belongs to the universality 
class of the RFIM, we expect, in the thermodynamic limit, a critical point at 
temperature $\tch$ and chemical potential $\mucr$. Below $\tch$, we anticipate 
bimodal distributions $P_{L,i}(N)$, but only if $\mu_i$ is chosen reasonably. In 
finite systems, we actually expect the bimodal form to persist considerably 
above $\tch$ also, since, for the RFIM, $P_{L,i}(N)$ remains sharp at 
criticality. In this work, $\mu_i$ is tuned for each realization of the porous 
medium such that $\partial \avg{N}_i / \partial \mu_i$ for that realization is 
maximized, with $\avg{N}_i = \sum_{N=0}^\infty N P_{L,i}(N)$ \footnote{Note that 
other choices are conceivable also, such as the \ahum{equal-area-rule} 
\cite{binder.landau:1984,borgs.kappler:1992}, or the generalized $k$-locus 
defined in \cite{orkoulas.fisher.ea:2001}; all become identical in the limit $L 
\to \infty$, of course.}. Loosely identifying $\avg{N}_i \leftrightarrow 
\avg{s}$, $\mu_i \leftrightarrow H$, our choice may be regarded as the analogue 
of the inflection point~A in \fig{fig3}(b) for the magnetic case.

\fig{fig4} shows a schematic of $P_{L,i}(N)$ in bimodal form. The left peak 
represents the vapor, the right peak the liquid, with the average $\avg{N}_i$ of 
the full distribution located in between (arrow A). \fig{fig4} also shows that 
$N_{\rm max}$ should be chosen well beyond the liquid peak (arrow B). If we 
shift $P_{L,i}(N)$ by its average, we approximately recover the Ising symmetry 
$P_L(s)=P_L(-s)$ of the magnetization distribution. Taking the quenched average, 
this requires a shift over $\qavg{\avg{N}} = (1/M) \sum_{i=1}^M \avg{N}_i$. 
Therefore,
\begin{equation}
 \label{eq:sub}
 \left( N - \qavg{\avg{N}} \right) / L^d,
\end{equation}
in a fluid with quenched disorder, is the analogue of~$s$ in a magnetic system, 
where the factor $L^d$ is needed because $s$ is the magnetization per spin. 
Replacing $s$ in the definitions of $\chi$ and $\chidis$ by \eq{eq:sub}, one 
obtains
\begin{eqnarray}
 \chi = \qavg{\avg{N^2} - \avg{N}^2} / L^d, \\
 \chidis = \left( \qavg{\avg{N}^2} - \qavg{\avg{N}}^2 \right) / L^d,
\end{eqnarray}
as the analogues of the connected and disconnected susceptibility in a fluid 
with quenched disorder. As stated before, $\chi$ and $\chidis$ are analyzed for 
the vapor and liquid phase separately, using $\avg{N}_i$ as a \ahum{cut-off} 
separating the peaks in $P_{L,i}(N)$. In this way, we obtain for the vapor phase
\begin{equation}
 \label{eq:nki}
 \avg{N^k}_{{\rm vap},i} = 2 \sum_{N=0}^{\avg{N}_i} N^k P_{L,i}(N),
\end{equation}
where the factor-of-two is a consequence of the normalization of $P_{L,i}(N)$. 
The moments $\avg{N^k}_{{\rm liq},i}$ of the liquid are obtained similarly, with 
the summation from $N=\avg{N}_i$ to $N_{\rm max}$. The connected and 
disconnected susceptibilities of the vapor phase can now be written as
\begin{equation}
 \hspace{-5mm}
 \chi_{\rm con}^{\rm vap} = \frac{
 \qavg{\avg{N^2}_{\rm vap}} - \qavg{\avg{N}_{\rm vap}^2}}{L^d}, \hspace{5mm}
 \chi_{\rm dis}^{\rm vap} = \frac{
 \qavg{\avg{N}^2_{\rm vap}} - \qavg{\avg{N}_{\rm vap}}^2}{L^d},
\end{equation}
with the quenched average $\qavg{\cdot}$ conveniently expressed in terms of 
\eq{eq:nki} as $\qavg{\avg{N^k}_{\rm vap}^l} = (1/M) \sum_{i=1}^M 
\avg{N^k}_{{\rm vap},i}^l$. Similar expressions hold for $\chi_{\rm con}^{\rm 
liq}$ and $\chi_{\rm dis}^{\rm liq}$ also. Note that, since the chemical 
potential $\mu_i$ is \ahum{fine-tuned} for each realization of the porous 
medium, the quenched average obtained above actually corresponds to 
$\overline{X}$ of \eq{eq:ob}, but this should be obvious from our discussion. 
For completeness, we remark that the quenched-averaged distance between the 
peaks in \fig{fig4} may be used as order parameter $m = \qavg{\avg{N}_{\rm liq}} 
- \qavg{\avg{N}_{\rm vap}}$, although in this work the emphasis is on the 
susceptibilities.

Of course, it needs to be verified in simulations if the expected bimodal form 
of $P_{L,i}(N)$ really occurs in practice. In our previous work, this turned out 
to be the case \cite{vink.binder.ea:2006}. However, GC simulations of the 
Lennard-Jones fluid with quenched disorder have revealed distributions with 
three peaks also \cite{alvarez.levesque.ea:1999}; the possibility of two fluid 
phase transitions occurring has also been suggested \cite{page.monson:1996}, 
although this probably does not survive in the quenched average 
\cite{sarkisov.monson:2000}.

\section{Model and simulation method}
\label{method}

\subsection{AO model}

We now proceed to test the concepts of the previous section in a colloid-polymer 
mixture with quenched disorder. Our primary aim is to measure the connected and 
disconnected susceptibilities, and to show that {\it both} diverge at 
criticality. To describe the mixture, we use the AO model 
\cite{asakura.oosawa:1954, vrij:1976}. In this model, colloids (species~c) and 
polymers (species~p) are treated as spheres with respective diameters $\sigmac$ 
and $\sigmap$. Hard sphere interactions are assumed between colloid-colloid and 
colloid-polymer pairs, while the polymer-polymer interaction is taken to be 
ideal. In this work, $\sigmac$ is the unit of length, the colloid-to-polymer 
size ratio $q \equiv \sigmap/\sigmac=1$, and the spatial dimension will be 
$d=3$. The behavior of this model for $q=1$ {\it without} quenched disorder has 
been studied before \cite{fortini.schmidt.ea:2006}, and bulk phase separation, 
whereby the mixture \ahum{splits} into a colloid-rich (polymer poor) and 
colloid-poor (polymer rich) domain, was readily observed. If one ``identifies'' 
the colloid-rich phase with a liquid, and the colloid-poor phase with a vapor, 
the phase separation can be treated in much the same way as a \LV transition. In 
the GC ensemble, one then introduces the colloid chemical potential $\mu$, and, 
following convention, the polymer \ahum{chemical potential} 
$\etapr$~\footnote{Strictly speaking, $\etapr$ is defined as the polymer 
reservoir packing fraction \cite{lekkerkerker.poon.ea:1992}. For the present 
case of {\it ideal} polymers $\etapr = \pi \sigmap^3 e^{(\mu_{\rm p} / \kb T)} / 
6 \Lambda^3$, with $\mu_{\rm p}$ the polymer chemical potential, and $\Lambda$ 
the thermal wavelength.}. Phase separation occurs at the coexistence colloid 
chemical potential $\mu=\mucx$, for values of $\etapr$ exceeding the critical 
value $\etaprcr$ ($\etapr$ is therefore the analogue of inverse temperature; for 
$q=1$, $\etaprcr \approx 0.861$ has been reported 
\cite{fortini.schmidt.ea:2006}). The discussion and definitions of 
\sect{sec:fluid} thus trivially \ahum{carry-over} to the AO model if one 
identifies $N \leftrightarrow \mbox{number of colloids}$, $\mu \leftrightarrow 
\mbox{colloid chemical potential}$, and $T \leftrightarrow 1/\etapr$.

\subsection{AO model with quenched disorder}

To study the AO model with quenched disorder, we introduce a third species~Q of 
immobile (quenched) particles. These particles are also spheres, with diameter 
$\sigmaq=\sigmac$, and they are distributed in the simulation box at the start 
of each simulation (the simulation box, incidentally, is a cube of volume 
$V=L^d$ with periodic boundary conditions). The quenched particles, $N_{\rm Q}$ 
of them in total, are located at random positions, irrespective of overlap. 
Consequently, the structure of the quenched system is just that of an ideal gas. 
The average packing fraction of the quenched system is fixed at $\eta_{\rm Q} = 
\pi \sigmaq^3 N_{\rm Q} / (6V) = 0.05$, but, consistent with our GC approach, we 
allow for Poissonian fluctuations around the average. From a computational point 
of view, the quenched system is trivial to generate: one simply draws $N_{\rm 
Q}$ from a Poisson distribution, and generates a corresponding number of 
positions in the simulation box. Next, a GC simulation of the AO model is 
performed in the simulation box containing the quenched system, whereby the 
colloid and polymer positions are continuously updated, but not the positions of 
the quenched particles, of course. The colloids and polymers interact with the 
quenched particles in a simple way: colloids may {\it not} overlap with quenched 
particles, while the polymers may overlap {\it freely} with them. Of course, 
computational efficiency is the main motivation for using such simple 
interactions, although one could envision similar interactions in experiments 
also, using polymer quenched disorder. In any case, the simple approach adopted 
here is appealing, as previous work indicates 
\cite{schmidt.scholl-paschinger.ea:2002, vink.binder.ea:2006, 
pellicane.vink.ea:2008}. An estimate $\etaprcr \approx 1.192$ has also already 
been reported \cite{vink.binder.ea:2006}, for the exact same parameters as 
considered here.

\subsection{implementation details}

We now discuss some implementation details. For the $i$-th realization of the 
quenched system, grand canonical MC is used to measure $P_{L,i}(N)$ of 
\eq{eq:pn}, with $N$ the number of colloids. The distribution is obtained using 
the (already mentioned) biased sampling scheme \cite{virnau.muller:2004}, in 
conjunction with a cluster move \cite{vink.horbach:2004*1, vink:2004}. The 
cluster move is needed to alleviate the otherwise (too) slow equilibration of 
the AO model. Of course, simulations of a single-component fluid do not require 
the cluster move. We consider system sizes $L=7-12$. For each system size, 
$P_{L,i}(N)$ is typically measured for $M \sim 2000$~(!) realizations of 
quenched disorder, at several values of $\etapr$ in the vicinity of $\etaprcr$. 
Large values of $M$ are needed to obtain $\chidis$ accurately.

In GC simulations, particles are continuously inserted and deleted from the 
simulation box, and so one can define a time $\tau$ after which a given 
population of particles has been completely \ahum{updated} by new ones. The 
duration of a GC simulation may therefore be expressed in units of $\tau$. In 
the biased sampling scheme \cite{virnau.muller:2004}, simulation time can be 
conveniently allocated, since the scheme constructs $P_{L,i}(N)$ step-by-step 
via so-called windows. In the first window, $N$ varies between 0 and 1, in the 
next window between 1 and 2, and so forth, up to $N_{\rm max}$ (the number of 
polymers $N_{\rm p}$ fluctuates freely in each window, of course). Hence, we 
allocate a fixed amount of simulation time, typically $5\tau$, to each window. 
It then takes roughly 12~minutes to obtain $P_{L,i}(N)$ for $L=7$, and about 
1~hr for $L=12$. Of course, these benchmarks depend on $\etapr$, as well as on 
the precise computer architecture, but they suffice to give an overall 
impression of how much computer time was used.

A final remark concerns the implementation of histogram extrapolation 
\cite{ferrenberg.swendsen:1988}. As stated earlier, all simulations are 
performed at colloid chemical potential $\mu_i=0$, and $P(N|\mu_i=\mu') \propto 
P(N|\mu_i=0) \exp(\mu' N)$ is used to extrapolate to different values. 
Obviously, a similar expression holds for the polymers also, which one could use 
to extrapolate in $\etapr$. In fact, an important ingredient of this work is 
precisely the latter extrapolation, and our analysis would become extremely 
cumbersome without it. However, this requires that we store the full 
two-dimensional histogram $P_{L,i}(N,N_{\rm p})$, with $N$ the number of 
colloids, and $N_{\rm p}$ the number of polymers. Since we typically 
consider~2000 realizations of quenched disorder, storage requirements become 
enormous. Fortunately, storage can be reduced tremendously, when one realizes 
that, for a fixed number of colloids $N$, the corresponding distribution in 
$N_{\rm p}$ is to a good approximation a single Gaussian peak. For $N=0$ this is 
obvious, since then we have a pure polymer system, but it holds well for $N>0$ 
also. Hence, to facilitate extrapolations in $\etapr$, we only need to store the 
average and variance in $N_{\rm p}$ for each window (which costs only very 
little storage, at no cost in CPU time either). We have verified this approach 
and checked that results obtained at one value of $\etapr$ indeed extrapolate to 
those obtained at a different value (not too far away, of course). Note also 
that the histogram extrapolation method itself can be optimized since, for a 
Gaussian distribution, integrations over $N_{\rm p}$ can be performed exactly 
beforehand; the resulting expressions become functions of the average and 
variance, which can be hard-coded.

\section{Results}
\label{result}

\subsection{sample to sample fluctuations}

\begin{figure}
\begin{center}
\includegraphics[clip=,width=13cm]{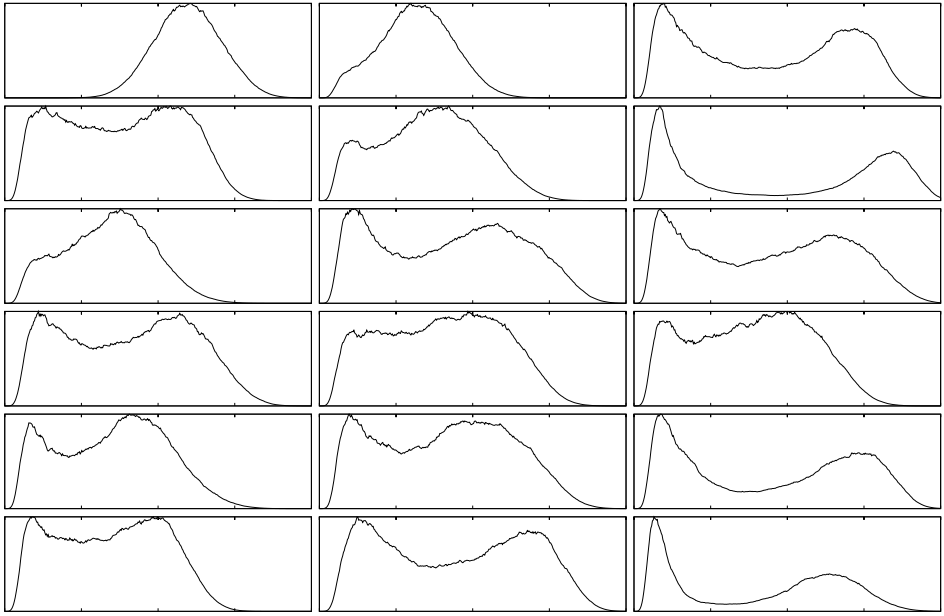}
\caption{\label{d105} Distributions $P_{L,i}(N)$ for 18 different realizations 
of quenched disorder using $\etapr=1.05$ and $L=10$. The horizontal axes in each 
of the plots show the colloid packing fraction $\etac \equiv \pi \sigmac^3 N / 
(6V)$ from $\etac=0 \to 0.2$ (left to right); the unit on the vertical axes is 
arbitrary.}
\end{center}
\end{figure}

\begin{figure}
\begin{center}
\includegraphics[clip=,width=13cm]{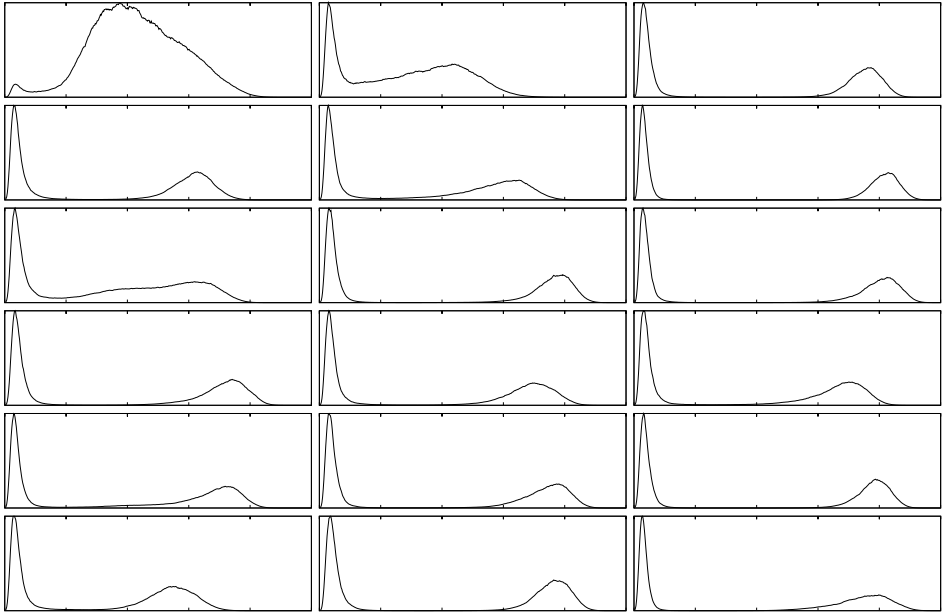}
\caption{\label{d115} Same as \fig{d105} but for $\etapr=1.15$ and $L=10$; the 
colloid packing fraction on the horizontal axes is from $\etac=0 \to 0.25$.}
\end{center}
\end{figure}

The analysis of \sect{sec:fluid} requires that the distributions $P_{L,i}(N)$ 
are somewhat bimodal, i.e.~that they resemble the schematic shape of \fig{fig4}. 
In order to verify this, we show, in \fig{d105}, $P_{L,i}(N)$ for a number of 
realizations of quenched disorder, at a value of $\etapr$ significantly {\it 
below} the critical value $\etaprcr \approx 1.192$. Clearly, the bimodal shape 
is already present in most distributions, even for this low value of $\etapr$. 
Of course, by making $\etapr$ even lower, the bimodal shape will eventually 
vanish for all realizations of quenched disorder, since we then enter the 
one-phase region where $P_{L,i}(N)$ is just a single peak, conform \fig{plm}(b). 
In any case, \fig{d105} does confirm our expectation that, for random-field 
Ising universality, bimodal distributions persist well above $\tch$ (recall that 
$\etapr$ is the analogue of inverse temperature). \fig{d105} also reveals that 
not all the distributions are bimodal, see, for example, the distribution in the 
upper left corner. In these cases, splitting the distribution in half at the 
average is not meaningful anymore, although numerically this can still be 
applied. Since, for the thousands of distributions generated in our simulations, 
inspecting each one visually by hand is not feasible, the (occasional) 
single-peaked distribution is treated in the same way as the bimodal ones. Of 
course, single-peaked distributions become increasingly rare upon increasing 
$\etapr$, as \fig{d115} clearly indicates, where the same realizations of 
quenched disorder were used as in \fig{d105}. Note that, in \fig{d115}, all 
distributions now feature two peaks.

Another feature that emerges from these figures is that the vapor peak is much 
sharper than the liquid peak. This appears to be a non-universal feature that 
depends on the interaction between fluid and quenched particles. In our previous 
work, we have studied a different type of quenched disorder, whereby also the 
polymers were not allowed to overlap with the quenched species 
\cite{vink.binder.ea:2006, pellicane.vink.ea:2008}. In this case, a reversed 
trend was observed, namely a sharp liquid peak \ahum{coexisting} with a much 
broader vapor.

\begin{figure}
\begin{center}
\includegraphics[clip=,width=13cm]{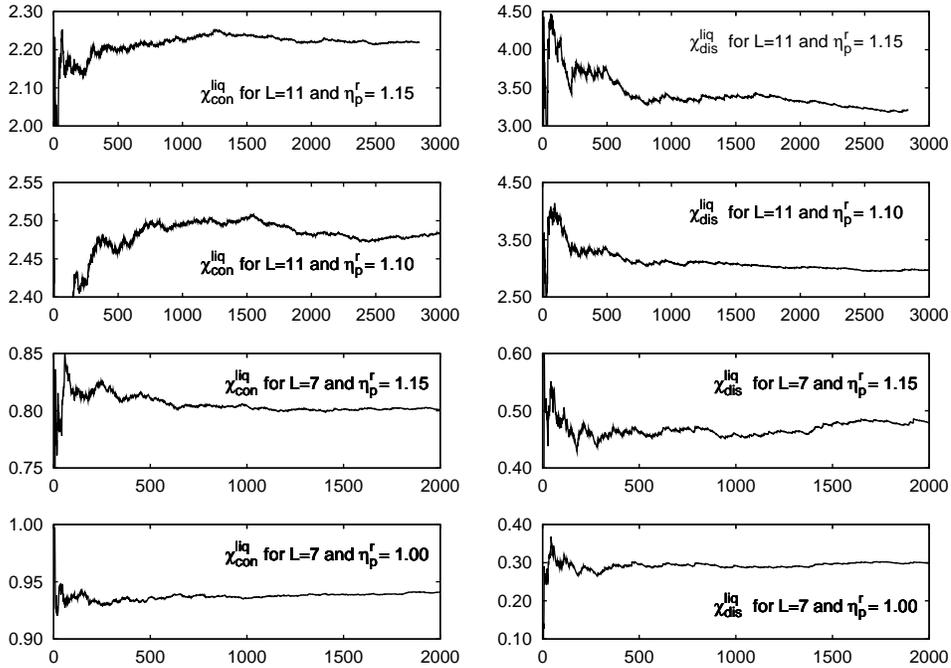}
\caption{\label{mavg} \ahum{Moving average} of the susceptibilities $\chi_{\rm 
con}^{\rm liq}$ (left frames) and $\chi_{\rm dis}^{\rm liq}$ (right frames) of 
the liquid phase, for $L=7$ and~$11$, using several values of $\etapr$ as 
indicated in the labels. Plotted are the susceptibilities (vertical axes) versus 
the number of quenched disorder realizations $M$ (horizontal axes).}
\end{center}
\end{figure}

Having shown that our assumption of bimodal distribution shape is a reasonable 
one, we need to determine the number of quenched disorder realizations $M$ 
typically required to obtain $\chi_{\rm con}^{\rm liq}$ and $\chi_{\rm dis}^{\rm 
liq}$ accurately. To this end we show, in \fig{mavg}, the variation of these 
quantities as a function of $M$, for two system sizes $L$, and several values of 
$\etapr$ (as indicated in the label of each subplot). The trends revealed in 
\fig{mavg} are typical for other state-points also. Clearly, from this figure, 
we conclude that $M$ should be of the order of 1000 at least. Larger values are 
better still, but then we meet the limit of our computational resources. 

\begin{figure}
\begin{center}
\includegraphics[clip=,width=8cm]{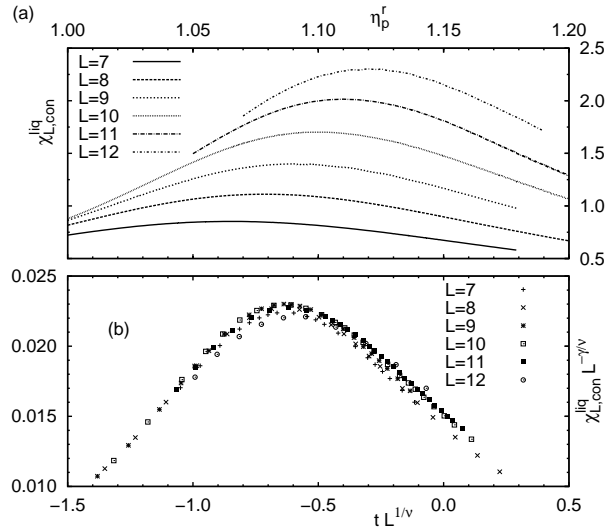}
\caption{\label{susl} (a) Connected susceptibility of the liquid phase versus 
$\etapr$ for several system sizes $L$. Note the increase of peak 
height with $L$, and also the shift in the peak positions. (b) Finite size 
scaling plot, where $\etaprcr=1.194$, $\nu=1.1$, and $\gamma/\nu=1.87$ were used 
(see details in text).}
\end{center}
\end{figure}

\begin{figure}
\begin{center}
\includegraphics[clip=,width=8cm]{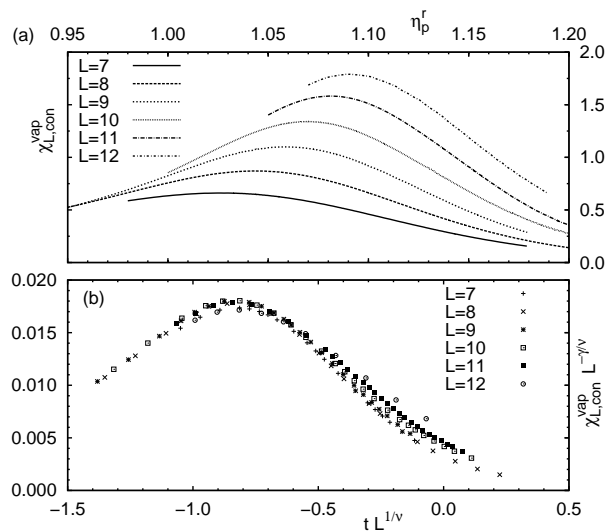}
\caption{\label{susv} Same as \fig{susl} but for the connected susceptibility of 
the vapor. In the scaling plot of (b), $\etaprcr=1.194$, $\nu=1.1$, and 
$\gamma/\nu=1.87$ were used.}
\end{center}
\end{figure}

\subsection{connected susceptibility}

We now consider the connected susceptibility, first of the liquid phase. Shown 
in \fig{susl}(a) is $\chi_{\rm con}^{\rm liq}$ versus $\etapr$, for several 
system sizes. Note the presence of the peak. Consistent with finite size 
scaling, the peak height increases with $L$; the latter could now be fitted to 
\eq{eq:fss:chi} to obtain $\gamma/\nu$. However, a more stringent test is to 
plot
\begin{equation}
 \label{eq:col}
 t L^{1/\nu} \hspace{3mm} \mbox{vs.} \hspace{3mm} 
 \chi_{L, \rm con}^{\rm liq} L^{-\gamma/\nu}, 
\end{equation}
with $t = \etapr/\etaprcr-1$ the relative distance from the critical point. 
Although not derived in this work, finite size scaling implies that data from 
different system sizes, when scaled conform \eq{eq:col}, collapse onto a 
single master curve, provided the correct values of $\etaprcr$, $\nu$, and 
$\gamma$ are used~\footnote{The derivation is straightforward, see for example 
\cite{newman.barkema:1999}.}. In \fig{susl}(b) the resulting scaling plot is 
shown, where $\etaprcr=1.194$, $\nu=1.1$, and $\gamma/\nu=1.87$ were used. The 
quality of the collapse is clearly very good. However, we noticed that good 
collapses were obtained for different values also, typically $\nu=1.0-1.2$ and 
$\etaprcr=1.19-1.22$, which gives an indication of the uncertainty. The problem 
is that both $\nu$ and $\etaprcr$ follow from the $L$-dependence of the peak 
positions. Over the range of available system sizes, the shift in the peak 
positions is rather small, and hence large uncertainties in $\nu$ and $\etaprcr$ 
are unavoidable. In contrast, $\gamma/\nu$ can be obtained more reliably, since 
the latter is set by the peak height versus $L$, which yields a more pronounced 
numerical signature. Similar conclusions are reached for the connected 
susceptibility of the vapor, see \fig{susv}.

\begin{figure}
\begin{center}
\includegraphics[clip=,width=8cm]{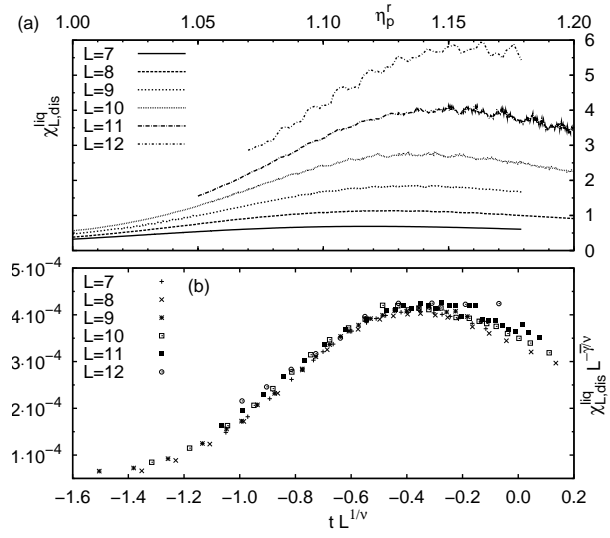}

\caption{\label{disl} The main result of this paper: (a) disconnected 
susceptibility of the liquid phase versus $\etapr$ for several system sizes $L$, 
and (b) the corresponding finite size scaling plot, where $\etaprcr=1.194$, 
$\nu=1.1$, and $\bar{\gamma}/\nu=3.82$ were used (see details in text).}

\end{center}
\end{figure}

\begin{figure}
\begin{center}
\includegraphics[clip=,width=8cm]{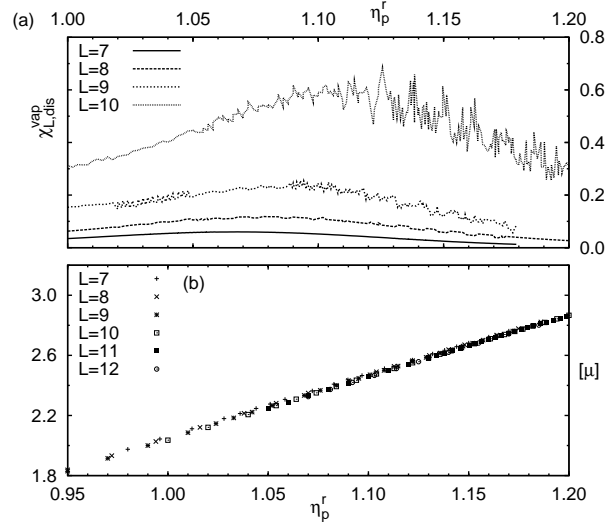}
\caption{\label{disv} (a) Disconnected susceptibility of the vapor phase versus 
$\etapr$, and (b) the quenched-averaged chemical potential versus $\etapr$ (for 
several system sizes $L$).}
\end{center}
\end{figure}

\subsection{disconnected susceptibility}

We now come to the main result of this work, namely the behavior of the 
disconnected susceptibility. If fluids with quenched disorder belong to the 
universality class of the RFIM, the analogue of $\chidis$ defined in 
\sect{sec:fluid} should diverge with critical exponent $\bar{\gamma}$. Since 
$\bar{\gamma} = 2\gamma$ \cite{schwartz:1985}, the divergence should be very 
pronounced, much more pronounced than that of the connected susceptibility, in 
fact. In \fig{disl}(a), we show $\chi_{\rm dis}^{\rm liq}$ of the liquid phase 
versus $\etapr$ for several system sizes. The formation of a peak is clearly 
visible. Note also the rapid growth of the peak height: increasing the system 
size from $L=7 \to 12$, the disconnected susceptibility increases by a factor of 
more than six, compared to a factor of about three for the connected 
susceptibility. The corresponding scaling plot is shown in \fig{disl}(b), which 
now involves $\bar{\gamma}$, of course. Using $\etaprcr=1.194$, $\nu=1.1$, and 
$\bar{\gamma}/\nu=3.82$, the data collapse convincingly, confirming the power 
law divergence of $\chidis$. For the same reason as before, the scaling plot is 
rather insensitive to $\etaprcr$ and $\nu$, and so the uncertainty in these 
quantities is similar as before, but the ratio $\bar{\gamma}/\nu$ should again 
prove reliable. In \fig{disv}(a) we plot the disconnected susceptibility of the 
vapor phase versus $\etapr$, but only for $L \leq 10$. For reasons we do not yet 
fully understand, the statistical uncertainty in $\chi_{\rm dis}^{\rm vap}$ is 
very large. While the growth of a peak with system size is still confirmed, the 
data clearly do not lend themselves for measuring critical exponents, and so a 
scaling plot is not attempted here. One reason for the large statistical 
uncertainty in $\chi_{\rm dis}^{\rm vap}$ is the smaller number of particles 
in the vapor phase (compared to the liquid). 

\begin{figure}
\begin{center}
\includegraphics[clip=,width=8cm]{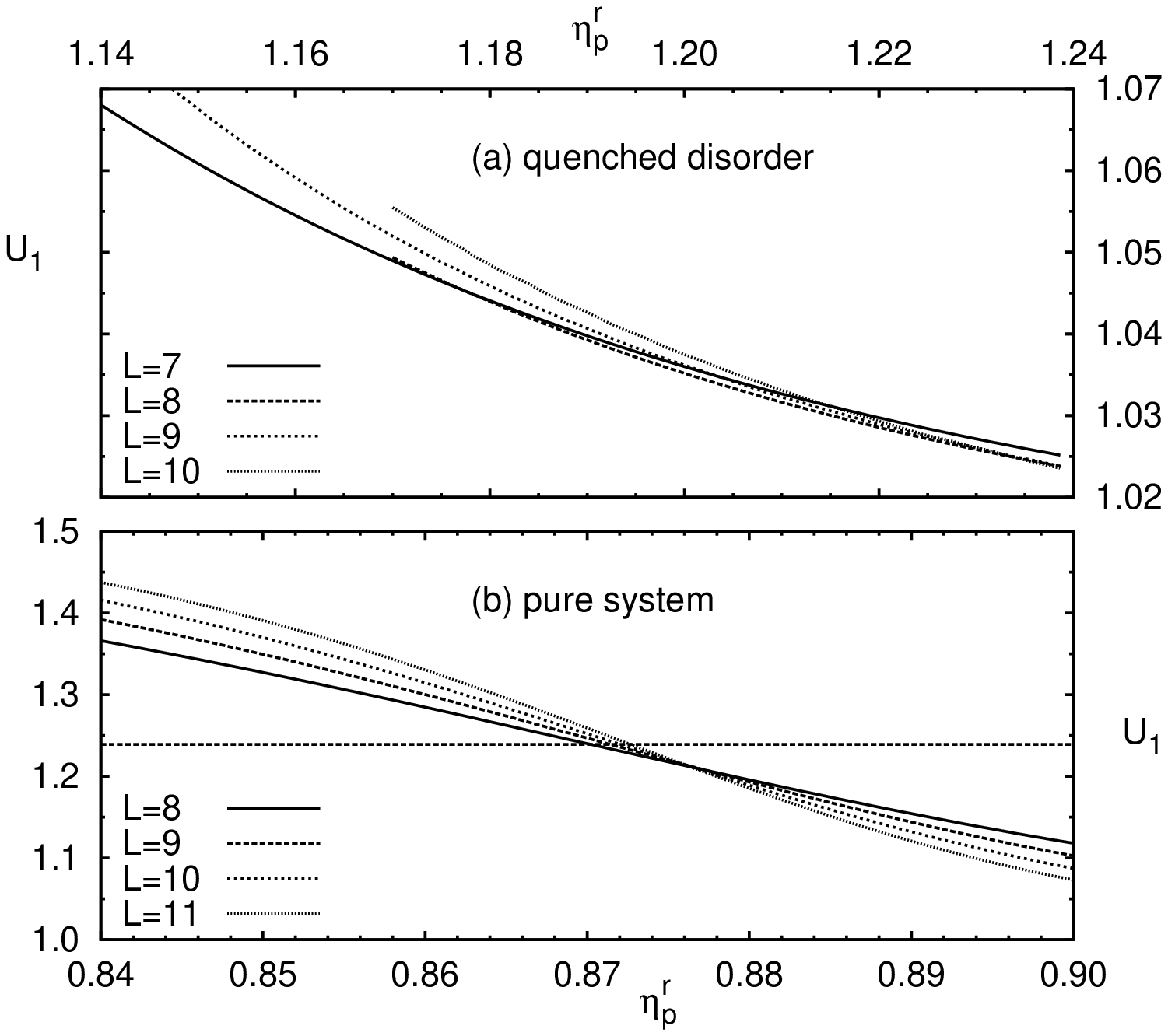}

\caption{\label{cumulant} Plots of $U_1$ versus $\etapr$, using several system 
sizes $L$, for the AO model with quenched disorder (a), and without (b). The 
horizontal line in (b) marks $U_1^\star \approx 1.239$ of the Ising model in 
three dimensions.}

\end{center}
\end{figure}

\setcounter{footnote}{1}

\subsection{scaling of the cumulant}

In \fig{cumulant}(a), we show the cumulant as a function of $\etapr$ for several 
system sizes. Recall that the cumulant is defined as $U_1 \equiv 
\qavg{\avg{s^2}} / \qavg{\avg{|s|}^2}$, with $s = \left(N - \qavg{\avg{N}} 
\right) / L^d$, which can be calculated straightforwardly from the distributions 
$P_{L,i}(N)$~\footnote{In our previous work \cite{vink.binder.ea:2006}, we used 
$U_1 \equiv \qavg{\avg{s^2}} / \qavg{\avg{|s|}}^2$, but the reader can verify 
following \sect{sec:fss_rfim} that both definitions become equivalent for $L \to 
\infty$.}. As expected, the cumulants from different system sizes do not 
intersect at criticality, but instead reveal a scatter of intersection points, 
close to $U_1^\star=1$ of the RFIM. This behavior is conform our discussion of 
\sect{sec:fss_rfim}, and confirms that $P_{L,i}(N)$ remains {\it sharp} at the 
critical point, featuring two well-separated peaks, since hyperscaling is now 
violated.

For comparison, \fig{cumulant}(b) shows the cumulant of the AO model in the pure 
system, i.e.~without quenched disorder. In this case, hyperscaling is not 
violated, and a sharp intersection point is indeed revealed, occurring at a value 
$U_1^\star$ {\it different} from the off-critical values $1$ and $\pi/2$, 
respectively. For Ising systems in $d=3$ dimensions, we expect that $U_1^\star 
\approx 1.239$ \cite{luijten.fisher.ea:2002}, marked by the horizontal line in 
\fig{cumulant}(b), and our data indeed intersect close to this value (some 
deviation is clearly apparent, but to account for this would require a 
field-mixing analysis \cite{bruce.wilding:1992, vink.horbach:2004}). From the 
intersection point, we also conclude $\etaprcr \approx 0.876$ for the pure 
system, which compares well to the estimate reported in 
\cite{fortini.schmidt.ea:2006}.

\begin{figure}
\begin{center}
\includegraphics[clip=,width=8cm]{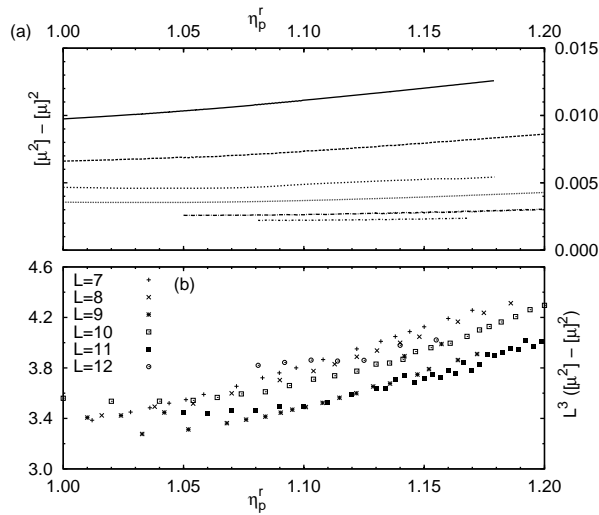}
\caption{\label{mhu} (a) Variance $\qavg{\mu^2} - \qavg{\mu}^2$ versus $\etapr$ 
for system sizes $L=7 \to 12$ (from top to bottom). (b) Same as above, but with 
the variance scaled by $L^d$.}
\end{center}
\end{figure}

\subsection{chemical potential}

Finally, we consider the average and variance of the chemical potentials at 
which our data were obtained. Recall that, for each realization~$i$ of quenched 
disorder, we use a \ahum{fine-tuned} chemical potential $\mu_i$, chosen at the 
maximum of $\partial \avg{N}_i / \partial \mu_i$ for that realization. Hence, it 
is interesting to consider the quenched-averaged chemical potential $\qavg{\mu}$ 
and its variance $\qavg{\mu^2} - \qavg{\mu}^2$, with $\qavg{\mu^k} = (1/M) 
\sum_{i=1}^M \mu_i^k$. Shown in \fig{disv}(b) is $\qavg{\mu}$ versus $\etapr$ 
for several system sizes. The data do not reveal any strong $L$-dependence, 
which is similar to that observed in fluids without quenched disorder. Of more 
interest is the variance, which should vanish for $L \to \infty$. Shown in 
\fig{mhu}(a) is $\qavg{\mu^2} - \qavg{\mu}^2$ versus $\etapr$, and the decrease 
of this quantity with increasing $L$ is clearly visible. Kierlik \etal have 
shown that, below $\tch$ in the two-phase region, the variance of the chemical 
potential vanishes $\propto L^{-d}$ \cite{kierlik.monson.ea:2002}. Plotting 
therefore $L^d \left(\qavg{\mu^2} - \qavg{\mu}^2 \right)$ versus $\etapr$, see 
\fig{mhu}(b), we observe that this prediction holds quite well for our data 
also.

\section{Discussion and summary}

We have explained finite size scaling in the random-field Ising model, and shown 
how this technique may be applied to a fluid with quenched disorder. We have 
also defined the analogue of the disconnected susceptibility $\chidis$ for the 
latter. If fluids with quenched disorder belong to the universality class of the 
random-field Ising model, as conjectured by de~Gennes \cite{gennes:1984}, 
$\chidis$ should diverge at criticality, and so our definition facilitates 
further tests of this conjecture. To perform one such test has been the topic of 
the present work, using the Asakura-Oosawa model of a colloid-polymer mixture 
confined to a random porous medium. Our data are indeed compatible with a 
divergence of $\chidis$. Moreover, for the liquid phase, we even recover 
$\bar{\gamma} \approx 2\gamma$, in quantitative agreement with the prediction of 
Schwartz for the random-field Ising model \cite{schwartz:1985}. Our estimate of 
the correlation length exponent $\nu \approx 1.0 - 1.2$, although not very 
precise, is also consistent with reported random-field Ising estimates 
\cite{rieger:1995, newman.barkema:1996}. Hence, the present results confirm our 
earlier study \cite{vink.binder.ea:2006}, where evidence of random-field Ising 
universality in fluids with quenched disorder was also presented, but based on 
the shape of $P_{L,i}(N)$ at criticality. We also remind the reader of the large 
number of porous medium realizations used in our analysis. As the \ahum{moving 
averages} of \fig{mavg} indicate, such numbers become a necessity, if $\chi$ and 
$\chidis$ are to be obtained with any meaningful accuracy.

Finally, we turn to a discussion of possible applications of our work to 
experiments. The prototype experimental realization of a fluid with quenched 
disorder is an {\it atomic} fluid injected into silica aerogel. This realization 
has the disadvantage that the coupling between the porous medium and the fluid 
is weak \cite{wong.chan:1990, wong.kim.ea:1993}, as manifested by the small 
shift of the critical temperature (compared to the system without quenched 
disorder). Moreover, the characteristic length over which the aerogel structure 
appears random is very large, compared to the size of the fluid molecules. In 
this respect, colloidal fluids may offer an attractive alternative. Note that 
investigations of critical phenomena in colloid-polymer mixtures {\it without} 
quenched disorder \cite{poon:2002, ramakrishnan.fuchs.ea:2002} are already 
experimentally feasible: critical interface and density fluctuations can be 
visualized directly \cite{aarts.schmidt.ea:2004, royal.aarts.ea:2006} using 
confocal microscopy \cite{vossen.van-der-horst.ea:2004}. In principle, such 
confocal experiments could be extended to include quenched disorder also. The 
generation and synthetization of quenched colloidal porous media has received 
considerable attention \cite{hoa.lu.ea:2006, physreve.61.626, cho.kim.ea:2008}. 
One could envision an experiment whereby a colloid-polymer mixture is injected 
into a rigid colloidal gel. Such gels could be formed using small nanoparticles 
which can grow into randomly branched networks at volume fractions of only a few 
percent \cite{cho.kim.ea:2008}. The size of these nanoparticles can be much 
smaller than the typical colloid or polymer diameter, and so one can easily 
reach the regime where the critical correlations of the colloid-polymer mixture 
average over the random structure of the gel. Another feasible realization would 
be to use a polymer blend containing nanoparticles of suitable size, such that 
the diffusion of these particles in the blend is small. The structure formed by 
the nanoparticles will then appear to be frozen (quenched) on the timescales 
needed for the critical correlations of the polymer blend to equilibrate. The 
latter could then be measured using, for example, light scattering~\footnote{We 
are indebted to G.~Fytas for a stimulating discussion of this point.}. In any 
case, we hope that the simulational efforts of the present work will stimulate 
experimental efforts also, in order to completely settle this longstanding 
problem.

\ack

This work was supported by the {\it Deutsche Forschungsgemeinschaft} under the 
SFB-TR6 (project sections A5 and D3) and the Emmy Noether program (VI 483/1-1).

\section*{References}
\bibliographystyle{simple}
\bibliography{index}

\end{document}